\documentclass{iopart}

\usepackage[utf8]{inputenc}

\usepackage[T1]{fontenc}

\usepackage{times}
\usepackage{eucal}
\usepackage{dsfont}
\usepackage{mathrsfs}
\usepackage{bm}

\usepackage{xcolor}
\usepackage[]{graphicx}
\usepackage[normalem]{ulem}
\usepackage{siunitx}

\usepackage{booktabs}
\usepackage[colorlinks,allcolors=blue]{hyperref}
\renewcommand{\vec}[1]{{\ensuremath{\bm{\mathrm{#1}}}}}

\renewcommand{\exp}[1]{\ensuremath{{\mathrm{e}^{#1}}}}
\newcommand{\ii}{{\ensuremath{\mathrm{i}}}}

\begin{document}

\title{Electronic structure and finite temperature magnetism of yttrium iron garnet}

\author{Joseph Barker$^{1,2}$, Dimitar Pashov$^{3}$, Jerome Jackson$^{4}$}
\address{$^1$School of Physics and Astronomy, University of Leeds, Leeds LS2 9JT, United Kingdom}
\address{$^2$Bragg Centre for Materials Research, University of Leeds, Leeds LS2 9JT, United Kingdom}
\address{$^3$Department of Physics, King's College London, Strand, London WC2R 2LS, United Kingdom}
\address{$^4$Scientific Computing Department, STFC Daresbury Laboratory, Warrington WA4 4AD, United Kingdom}
\ead{j.barker@leeds.ac.uk}

\begin{abstract}
Yttrium iron garnet is a complex ferrimagnetic insulator with 20 magnon modes which is used extensively in fundamental experimental studies of magnetisation dynamics.  As a transition metal oxide with moderate gap (\SI{2.8}{eV}), yttrium iron garnet requires a careful treatment of electronic correlation.  We have applied quasiparticle self-consistent $GW$ to provide a fully ab initio description of the electronic structure and resulting magnetic properties, including the parameterisation of a Heisenberg model for magnetic exchange interactions.  Subsequent spin dynamical modelling with quantum statistics extends our description to the magnon spectrum and thermodynamic properties such as the Curie temperature, finding favourable agreement with experimental measurements.  This work provides a snapshot of the state-of-the art in modelling of complex magnetic insulators.
\end{abstract}


\maketitle

\section{Introduction}

Yttrium iron garnet (Y$_{3}$Fe$_{5}$O$_{12}$ - YIG) is a remarkable magnetic insulator~\cite{cherepanov_saga_1993} because its ultralow magnetic damping facilitates fundamental investigations of spin-waves, transport and magnons.  
YIG has played a key role in advances such as the  Bose-Einstein condensation of magnons~\cite{demokritov_boseeinstein_2006}, magnonics~\cite{serga_yig_2010}, spincaloritronics~\cite{bauer_spin_2012}, long distance spin transport in insulators~\cite{cornelissen_long-distance_2015} and superconducting magnetic qubits~\cite{tabuchi_coherent_2015}.  Although often treated simply as a ferromagnet, it is a ferrimagnet with a large unit cell, containing 8 Fe atoms at octahedral sites magnetically anti-aligned with 12 Fe atoms in tetrahedral environments; both have a formal $3+$ charge state~\cite{geller_crystal_1957}.  YIG is a popular base for doping as yttrium is easily substituted, partially or in full, by other rare-earth elements.  Doping with Bi is also common, as this increases the strength of magneto-optical effects.  Fe can be substituted by other magnetic atoms such as Co or non-magnetic elements such as Al and Ga, allowing some tuning of the magnetocrystalline anisotropy and the magnetisation compensation point.

The basic properties of YIG were studied extensively 50 years ago including the deduction of Heisenberg exchange interactions from inelastic neutron scattering measurements, although at that time only three of the twenty magnon modes were measured~\cite{plant_spinwave_1977}.  Recently, the complete magnon spectrum has received interest in an effort to understand the role of the higher energy magnons in thermal spin transport, such as in the spin Seebeck effect.  Contemporary neutron measurements have provided more detail on the magnon modes~\cite{princep_full_2017} and their polarisation~\cite{nambu_observation_2020}.

Conspicuously absent in the literature are electronic structure studies of YIG.  This is because YIG, as a ferrimagnetic oxide with moderate band gap, requires treatment of electronic correlation better than the methods typically used in density functional theory (local or semi-local exchange correlation functionals such as the LDA), while the 80-atom unit cell is sufficiently complex that the use of improved methods is technically challenging.  While GGA calculations provide insight into the distribution of spin in YIG~\cite{bouguerra_y3fe5o12_2007}, most progress has been made with +U: Nakamoto et al.\ provide a systematic study of the series of rare-earth iron garnets~\cite{nakamoto_properties_2017} and Xie et al.~\cite{xie_first-principles_2017} used DFT+U to fit a Heisenberg model using total energy differences for different spin configurations, with U chosen such that the resulting Curie temperature of the classical Heisenberg model agreed with the experimentally measured value.  The obtained magnon spectrum was similar to the historic neutron measurements, although the optical modes were too low in frequency.  The U found to give the correct Curie temperature shows a band gap (\SI{1.6}{eV}) significantly lower than measured in experiments (\SI{2.8}{eV}) and choosing U to fit $T_{\rm C}$ precludes a truly ab initio calculation of the Heisenberg exchange. We are unaware of any $GW$ or hybrid functional calculation of YIG.

In this work we will address several of these shortcomings.  Using quasiparticle self-consistent \textit{GW}~\cite{faleev_all-electron_2004,vanschilfgaarde_quasiparticle_2006} (QSGW), we provide an accurate ab initio description of the electronic structure that is capable of describing both delocalised bands and strongly localised states (such as the Fe \textit{d}-electrons in this case) simultaneously.  The screened Coulomb interaction ($W$) describes explicitly what is typically modelled by adding the Hubbard U in LDA+U calculations, without requiring the identification of a local strongly correlated subspace, or of choosing interaction parameters for that subspace (or double counting correction schemes).  $GW$ gives meaning to band structures in terms of ionisation and absorption energies and thereby solves the {\emph{band gap problem}} typical of local and semi-local methods in density functional theory.  $GW$ is, however, a perturbative method and the results depend upon the starting point -- conventionally for materials studies with $GW$ this is the LDA or GGA.  Quasiparticle self-consistent $GW$ is a method to achieve an optimal starting point by forming from the $GW$ self energy a static (but non-local) potential that replaces the original exchange-correlation potential.  This process is iterated until self-consistency, at which point the bands of the {\emph{quasiparticlised}} self energy coincide with the poles of the $GW$ spectral function; the resulting QSGW electronic structure can differ significantly from the LDA one (or the first iteration: one-shot $G^0W^0$, which is commonly referred to as $GW$).  The QSGW potential is a simpler object than the $GW$ self energy and this facilitates investigating many properties; here we use the QSGW potential to calculate the transverse spin susceptibility from which, by inversion, we parameterise a Heisenberg model of the magnetic interactions.

We study this model Hamiltonian using spin dynamics to calculate the magnon spectrum and thermodynamic properties such as the Curie temperature.  For solving the spin model, we adopt a more sophisticated method than is typically applied to materials studies of this kind by the use of a quantum thermostat~\cite{barker_semiquantum_2019}.  Classical thermal statistics over populates high energy modes resulting in incorrect thermodynamics~\cite{kuzmin_shape_2005}.  We show that in this case good agreement with experiment can only be achieved by using quantum thermal statistics.  Combining QSGW electronic structure calculations with rigorous treatment of the resulting spin model, this work is a snapshot of the current state-of-the-art of modelling of complex magnetic materials in terms of the sophistication of the methods and their accuracy.

\section{Methods}

\subsection{Electronic structure}
QSGW calculations were performed using the full-potential linearised muffin-tin orbital (FP-LMTO) code Questaal~\cite{pashov_questaal_2020} for the experimental cubic garnet structure with $Ia\overline{3}d$ symmetry~\cite{euler_oxygen_1965}, comprising 80 atoms.
The main task in QSGW involves determining the self energy $\Sigma = iGW$, which is implemented numerically using a sum-over-states evaluation of the $\omega$-dependent single-particle polarisability, $\chi^0$, which defines the screened Coulomb interaction $W$, followed by the calculation of $\Sigma(\omega)$ using the contour deformation technique;
a description of these methods (including and the treatment of the Coulomb singularity for $\vec{G}{=}\vec{G'}\rightarrow 0$) is presented in~\cite{kotani_quasiparticle_2007}.
Once the self energy has been calculated, the static QSGW potential is represented in the basis of single-particle eigenfunctions $\psi_i$ (with eigenvalues $\varepsilon_i$),
\begin{equation}
V^{QSGW} = \frac{1}{2}\sum_{ij} |\psi_i\rangle \left\{ {{\rm Re}[\Sigma({\varepsilon_i})]_{ij}+{\rm Re}[\Sigma({\varepsilon_j})]_{ij}} \right\} \langle\psi_j|
\label{eq:vqsgw}
\end{equation}
Note that this involves the calculation of the full $\Sigma$ matrix, including its off-diagonal elements, as a function of $\omega$.

$V^{QSGW}$ can be rotated in to the local basis, which has finite real-space range, allowing the interpolation of $\Sigma$ to $q$ points other than those for which the self energy is explicitly calculated;
this is important because the single-particle part of a QSGW calculation typically requires a denser sampling of the Brillouin zone (to accurately describe the kinetic energy) than is necessary to describe the $q$-dependence of the self energy.
Converged results are obtained for single-particle $k$ meshes with $8^3$ points, while the QSGW self energy is evaluated at $4^3$ {$q$} points (differences in the QSGW gap between $3^3$ and $4^3$ points are $<$\SI{1}{meV}).

The FP-LMTO setup is defined by almost touching augmentation spheres ($R{=}2.15, 1.95$ {a.u.} for Fe$_a$ and Fe$_d$, $2.65$ {a.u} for Y and $1.55$ {a.u.} for O) and (2-$\kappa$) smoothed Hankel functions up to $\ell_{\rm max}{=}2$.
Augmentation up to $\ell_{\rm max}^{a}{=}3$ is used, and this is also the limit used for truncation of wave function products (the product basis contains combinations of products of augmentation functions and interstitial plane wave parts) used in the $GW$ calculation.
High lying local orbitals corresponding to the Fe 4\textit{d} state are included.
Smoothed Hankel basis definitions are chosen to yield spatially well localised functions which makes the interpolation of $\Sigma$ more reliable: \SI{-2.0}{Ryd} is used for Hankel function energies, with smoothing radii chosen by an automatic procedure (see~\cite{pashov_questaal_2020});
we estimate the precision of this configuration to be ${\sim}\SI{10}{meV}$ in calculated band gaps, evaluated at the LDA level, decreasing to ${\sim}\SI{100}{meV}$ in QSGW due to some loss of precision in describing empty states.
The QSGW calculations are based on the collinear ferrimagnetic LDA ground state solution.
All calculations are scalar-relativistic, the spin-orbit interaction is ignored.

The QSGW calculation was attempted in 2018 and was found to be nearly infeasible on the available hardware due to excessive memory, IO and computational requirements.
Algorithmic improvements in the evaluation of $\chi^0$ and $\Sigma$ avoided all but a small fraction of the IO.
Together with a more flexible memory management these allowed efficient parallelisation across multiple levels of processes and threads, utilising a number of multi-GPU accelerated nodes.
The effort culminated with the ability to obtain the single cycle self energy in under 10 minutes on an Nvidia DGX-1 node with $8\times$V100 cards.
Typically convergence of the QSGW potential to RMS change \SI{1e-5} is achieved in $<15$ iterations.
Due to queuing system restrictions the present results were performed on 16 Marconi100 nodes totalling 64 Nvidia V100 GPUs, reaching peak performance of 425 TFLOPs.

\subsection{Heisenberg exchange parameters}

Our description of magnetic interactions is based on the non-interacting transverse spin susceptibility, $\chi^{0+-}(q,\omega)$, which is calculated similarly to the charge susceptibility used in the main $GW$ calculation, following the formalism developed by Kotani and van Schilfgaarde~\cite{kotani_spin_2008}, which was shown to give an extremely accurate description of the magnetism in the materials NiO and MnO, which are essentially similar to YIG.  We define an extended Heisenberg model using the inverse of the static transverse spin susceptibility calculated from both the LDA and QSGW band structures.  We limit our attention to isotropic interactions between Fe sites.

The main feature of the non-interacting transverse susceptibility is Stoner excitations; the spin wave spectrum manifests itself only in the full, interacting susceptibility.  In reference~\cite{kotani_spin_2008}, some reasonable approximations were made in order to calculate the full susceptibility via $[\chi^{+-}(q)]^{-1} = [\chi^{0+-}(q)]^{-1} - I$.  However, the full susceptibility is not required for the evaluation of a Heisenberg model.

The transverse spin susceptibility is simplified by defining a normalised magnetisation function that is nonzero only within the augmentation sphere of each specific site (the volume $R_i$) that describes the $\ell{=}0$ component of the spin density: $m_i(r)\propto m(r)/\int_{R_i} m(r)dr$,
\begin{equation}
\tilde\chi_{ij}^{0+-}(q,\omega) = \int_{R_i} \int_{R_j}  m_i(r) \chi^{0+-}(r,r',q,\omega)  m_j(r')\, d^3r d^3r'
\label{eq:chipm}
\end{equation}
using this quantity the Heisenberg parameters follow directly:
\begin{equation}
J_{ij}^{0}(q) = [\tilde\chi^{0+-}(q,\omega{=}0)]^{-1}_{ij}
\label{eq:x0jij}
\end{equation}
Of course a similar definition can be written down for $J$, as the inverse of the full susceptibility: evaluated at the spin wave energies $\omega_{SW}(q)$, the inverse of $\chi^{+-}(q,\omega_{SW})$ corresponds to the {\emph{renormalised}} $J$~\cite{katsnelson_magnetic_2004}, which reproduces the spin wave spectrum by construction but has a less transparent connection to the Heisenberg model than does the bare, static interaction.  Nevertheless, in the case that $I$ is local, which is reasonable for well localised moments, then the spin-wave dispersion given by the static limit of the full susceptibility, $\propto J(q,\omega{=}0)-J(0,\omega{=}0)$ is the same as that of the bare susceptibility, and in the following we investigate only the bare $J^0$, sidestepping the evaluation of the full susceptibility (for background, see Antropov's discussion~\cite{antropov_exchange_2003}).  We simplify the notation to $J$ and adopt the following definition of spin Hamiltonian
\begin{equation}
\mathscr{H} = -\frac{1}{2} \sum_{ij}J_{ij}\vec{S}_i \cdot\vec{S}_j
\label{eq:ham}
\end{equation}
where $\vec{S}_i$ is a classical spin vector of unit length.  Very often, the {\emph{magnetic force theorem}} (of Liechtenstein, Katsnelson, Antropov and Gubanov~\cite{liechtenstein_local_1987}) is used and important recent extensions of this theory to methods for treating strong Coulomb correlation in real materials include QSGW by Yoon et al.~\cite{yoon_magnetic_2019}, which allows a detailed decomposition of the exchange couplings into states of different character, and dynamical mean-field theory by Kvashnin and coworkers~\cite{kvashnin_exchange_2015}, which allows the treatment of materials where the electronic structure is significantly multi-determinantal.  The magnetic force theorem is related to the inverse susceptibility by Taylor expanding in the small quantity $(\chi^{0+-}(q)-\chi^{0+-}(0))/\chi^{0+-}(0)$~\cite{kotani_spin_2008}; it remains not obvious which method is to be preferred since the force theorem has the advantage of being derivable in terms of energy changes under infinitesimal spin rotations.  
Susceptibility calculations are performed on a mesh of 64 $q$-points.

\subsection{Atomistic spin dynamics}

To perform thermodynamic calculations based on equation~(\ref{eq:ham}) we simulate dynamics using the Landau-Lifshitz equation of motion
\begin{equation}
  \frac{\partial \vec{S}_i}{\partial t
} = - \gamma \left(\vec{S}_i \times \vec{H}_i + \eta \vec{S}_i \times \left( \vec{S}_i \times \vec{H}_i \right) \right)\end{equation}
where $\gamma = g\mu_{\rm B} / \hbar$ is the gyromagnetic ratio and $\eta$ is a damping constant.  The local field felt by each atomic moment is
\begin{equation}
  \vec{H}_i = \vec{\xi}_i - \frac{1}{\mu_{i}}\frac{\partial \mathscr{H}}{\partial \vec{S}_i}
\end{equation}
where the $\vec{\xi}_i$ are the stochastic processes of a Langevin thermostat.  
As a classical model, the thermostat is conventionally defined to give classical (Rayleigh-Jeans) statistics and $\vec{\xi}_i$ are white noise processes satisfying the fluctuation-dissipation relations $\langle \xi_{i\alpha}(t)\rangle = 0 $ and $\langle \xi_{i\alpha}(t) \xi_{j\beta}(t') \rangle  = 2\eta k_B T\delta_{ij}\delta_{\alpha\beta}\delta(t-t')/\gamma$, where $\alpha,\beta\in\{x,y,z\}$.  
However, the use of classical statistics is poorly justified, because the highest magnon modes in YIG (and many magnetic materials) are around $\hbar\omega/k_{\rm B} \approx \SI{1000}{K}$.  
Classical results are known not to reproduce fundamental low temperature results such as Bloch's law~\cite{kuzmin_shape_2005}. 
Recently, attempts have been made to address this by rescaling the magnon density of states either as a renormalisation of the temperature~\cite{Woo2015} based on an assumed dispersion or as a slightly complex post process once the dispersion has been calculated~\cite{Bergqvist2018}. 
Here, we use a simple but general approach of a thermostat which obeys the quantum fluctuation-dissipation theorem for magnons 
\begin{equation}
  \langle\xi_{i\alpha}(t)\rangle = 0; \quad \langle \xi_{i\alpha} \xi_{j\beta} \rangle_{\omega} = \frac{2\eta\delta_{ij}\delta_{\alpha\beta}}{\gamma}\left(\frac{\hbar \omega}{\exp{\hbar\omega / k_{\rm B} T} - 1}\right)
  \label{eq:noise}
\end{equation}
which produces Planck statistics and gives quantitatively correct thermodynamic calculations~\cite{barker_semiquantum_2019, ito_spin_2019}. 
Processes satisfying the frequency spectrum of Eq.~(\ref{eq:noise}) are simulated in real time from a set of coupled stochastic differential equations (details are given in references~\cite{Dammak2009} and \cite{barker_semiquantum_2019}).
The spin dynamics calculations use large supercells ($16^3$ for thermodynamics and $40^3$ unit cells for spectra). After an equilibration period (\SIrange{0.05}{0.1}{ns}), thermodynamic quantities such as the magnetisation, susceptibility and magnon spectrum are calculated from trajectories of at least \SI{0.1}{ns}. The simulated total magnetisation is calculated as $\vec{m}(T) = \langle\sum_{i}\mu_{i}\vec{S}_i\rangle_{T} / \sum_{i} \mu_{i}\vec{S}_i^{0}$, where $\vec{S}_{i}^{0}$ is the ground state of spin $i$ and $\langle \cdots \rangle_{T}$ denotes a thermodynamic average and the susceptibility is  $\chi(T) = \left(\langle m^2 \rangle - \langle m \rangle^2 \right) / k_B T$. We calculate the magnon spectrum by 
\begin{equation}
  \mathscr{S}^{\alpha\beta}(\vec{q}, \omega) = \sum_d \sum_{ij} \exp{-\ii\vec{q}\cdot(\vec{r}-\vec{r}')} \int_{-\infty}^{\infty} {\rm d}t \left\langle s_{i,d}^{\alpha}(\vec{r}, t) s_{j,d}^{\beta}(\vec{r}', t') \right\rangle_{T}
  \label{eq:spectrum}
\end{equation}
where $\langle\cdots\rangle_{T}$ is a correlation function at temperature $T$, $s^{\alpha}(\vec{r},t) = W^{\alpha\beta} S^{\beta}(\vec{r},t)$ transforms a spin to the reference frame of its ground state, $d$ labels each position in the unit cell.
This differs from the neutron scattering cross section by avoiding the magnetic structure factor, thus allowing us to calculate all modes in a reduced zone scheme. 
As a supercell method, our $q$-space resolution is limited by the number of unit cell repeats in the supercell.
Equation~(\ref{eq:spectrum}) is essentially the transverse spin susceptibility $\chi_{ij}^{+-}(q,\omega)$, but derived from the spin model, instead from \SI{0}{K} band theory which is used to calculate $\tilde{\chi}^{0+-}$  (defining $J$), and includes the effect of temperature and magnon-magnon interactions.
In principle this allows the calculation of properties such as the temperature dependence of: magnon lifetimes, magnon frequencies~\cite{Barker2016b} and spin wave stiffness~\cite{Tomasello2018a}. 
In ferrimagnets the low-$q$ dispersion can also change with temperature, for example becoming linear close to a magnetisation compensation point~\cite{Barker2013}.

\section{Results and Discussion}

\subsection{Electronic density of states}
\begin{figure}
\includegraphics[width=\textwidth]{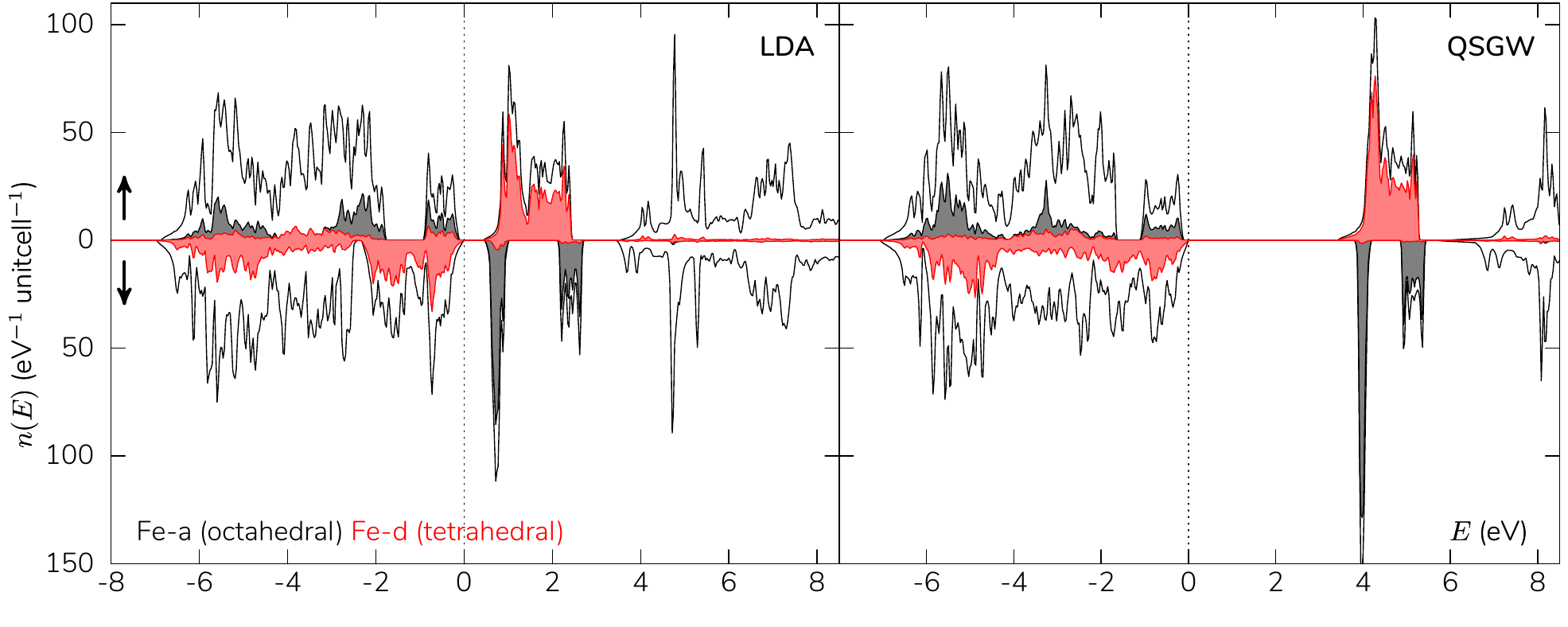}
\caption{Density of states of YIG calculated using spin-polarised LDA (left) or QSGW (right).  The partial density of states for octahedral Fe sites is shaded solid grey, that for tetrahedral Fe sites is coloured red and the total $n(E)$ is shown by the outline.  Upper and lower parts show spin-up and spin-down.  Zero of the energy range is set to the valance band maximum.
\label{fig:dos}}
\end{figure}

The LDA and QSGW densities of states are shown in Fig.~\ref{fig:dos}, where the contributions from octahedral and tetrahedral Fe sites are distinguished.  Because the Fe $d$-shells are half filled, fluctuations in the $d$ occupancy are unimportant and it is appropriate to describe the electronic structure using single-particle methods, such as both the LDA and the quasiparticlised QSGW potential; their general description of YIG as a ferrimagnetic band insulator is broadly similar.  QSGW principally corrects an underestimate in LDA of the exchange splitting on Fe sites, which drives an opening of the band gap, but also causes other changes throughout the valance and conduction bands.

The band gap in the LDA description (\SI{0.45}{eV}) increases in QSGW to \SI{3.4}{eV}.  This is larger than the experimental gap (\SIrange{2.8}{3.15}{eV}~\cite{wettling_optical_1973,metselaar_high-temperature_1974,wittekoek_magneto-optic_1975}) as is typical in QSGW and attributable to approximating the full susceptibility by the independent particle random phase approximation (RPA) one; this underestimates screening and yields a slightly too {\emph{Hartree-Fock-like}} description.  Including many-body contributions to the screening via the Bethe-Salpeter equation (BSE) effectively remedies this, but our BSE implementation does not yet scale to structures of this size.

As in the (more complicated) case of magnetite~\cite{antonov_electronic_2001}, Fe in the octahedral coordination (the Fe$_a$ sites) shows a large crystal field splitting and $e_g$ and $t_{2g}$ bands are separated by a gap in both LDA and QSGW; the splitting at tetrahedral sites is smaller and the bands overlap.  Compared to LDA, the crystal field splittings are reduced in QSGW from ${\sim}\SI{2}{eV}$ to ${\sim}\SI{1}{eV}$.

Concomitant with the increased exchange splitting, QSGW shows an increase in the Fe magnetic moments, resulting in $\mu_{a} = 4.17 \mu_B$ and $\mu_{d} = 3.93 \mu_B$ (see table~\ref{tab:comparison}).
This is naturally lower than the free ion value for Fe$^{3+}$ ($\mathcal{S}{=}5/2 {\sim} 5 \mu_B $), because of bonding and hybridisation in the crystal, and agrees very well with neutron diffraction measurements (although some uncertainty exists because small distortion to a $R\bar{3}$ symmetry structure occurs under specific applied fields~\cite{shamoto_neutron_2018}).
We find that QSGW generally yields moments in transition metal oxides that agree closely with experiment.

The magnitude of the total magnetisation of the unit cell calculated from first principles is identically $20 \mu_B$ in both LDA and QSGW; the part of the magnetisation not accounted for by Fe is found principally at O sites ($\mu{\sim} 0.07\mu_B$) which align with Fe$_d$, since the Fe-O bond is shorter at the tetrahedral sites.  The remainder resides in the  interstitial; yttrium, which occupies the $c$ positions in the garnet crystal, plays no role in the magnetism.  

\subsection{Exchange interactions and magnon spectrum}

\begin{figure}
\includegraphics[width=0.39\textwidth]{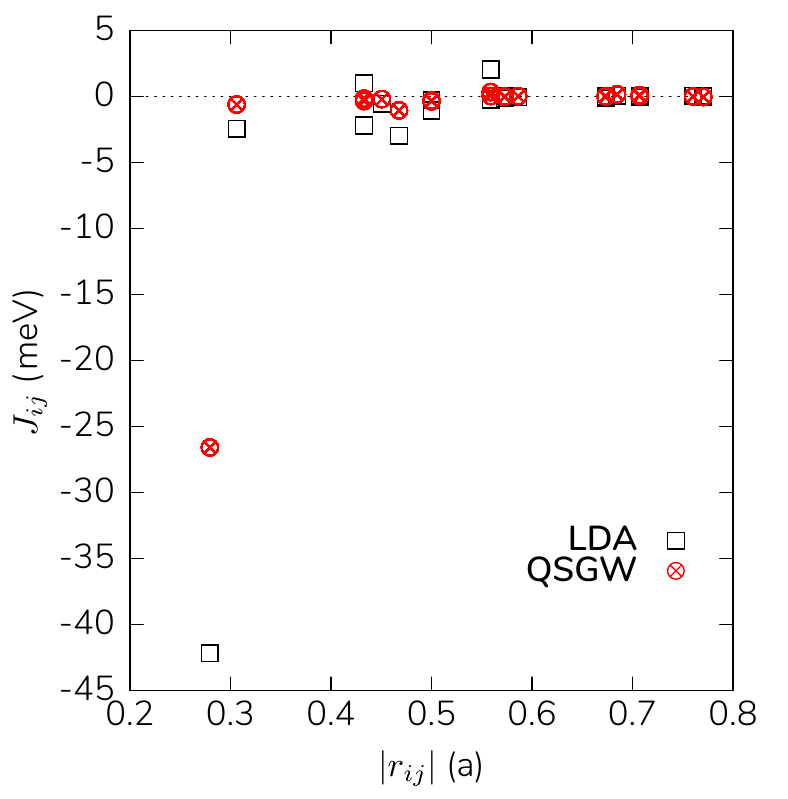}
\includegraphics[width=0.58\textwidth]{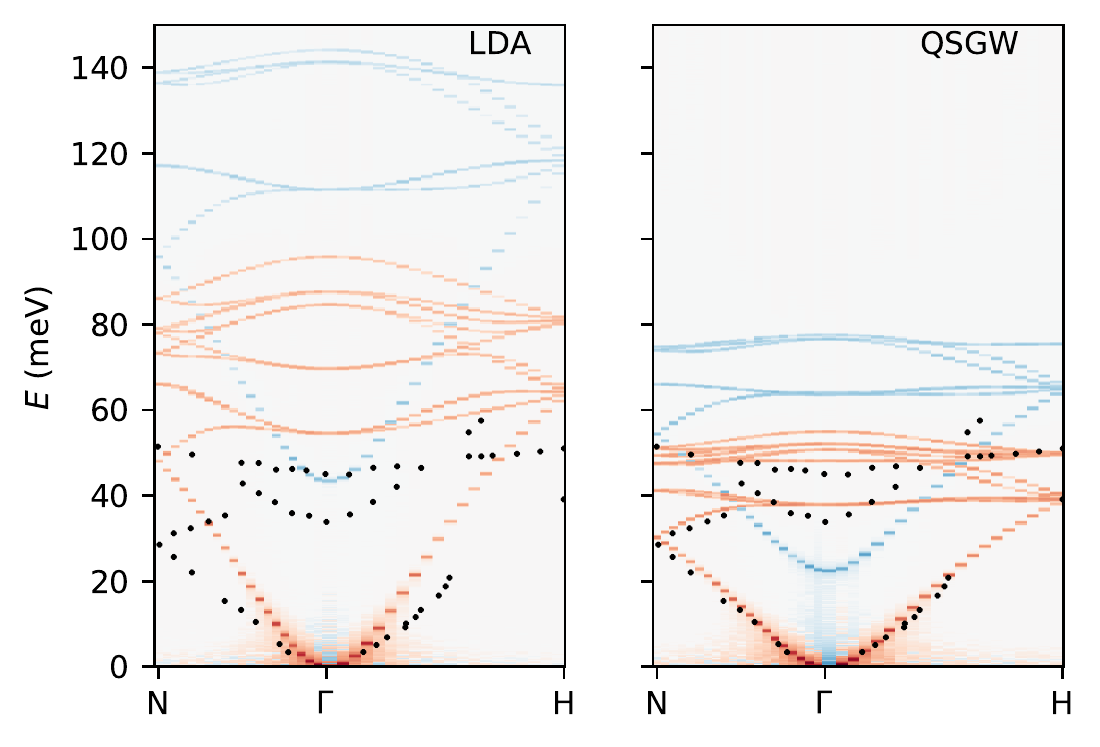}
\caption{Left: Comparison of LDA and QSGW calculated Heisenberg exchange constants. Right: Magnon spectral function $\mathscr{S}^{xy}(\vec{q}, \omega) - \mathscr{S}^{yx}(\vec{q}, \omega)$ along high symmetry lines in the BCC primitive cell.  Calculated from spin dynamics with $T=$\SI{10}{K}, $\eta=$\SI{2e-4}{}.  Red and blue modes have anti-clockwise and clockwise polarisations respectively. Black dots are experimental data from Ref.~\cite{plant_spinwave_1977}.
\label{fig:magnon_spectrum}}
\end{figure}

Fig.~\ref{fig:magnon_spectrum} shows the isotropic Heisenberg exchange interactions as a function of separation obtained from the LDA and QSGW susceptibilities.  The larger band gap in QSGW reduces the magnitude of the nearest neighbour superexchange interaction, as is well known.  Compared to the LDA model, the interactions in QSGW are shorter ranged and rapidly become negligible.  Our model does not exactly agree with parameters fitted to neutron scattering spectra (the most recent work by Princep et al.~\cite{princep_full_2017} reports the large nearest neighbour interaction to be \SI{-42}{meV}~\cite{princep_full_2017}, in the convention of Eq.~(\ref{eq:ham}), while we obtain \SI{-26.6}{meV}); we note that there is some variation of fitted $J$s and we compare instead the spin wave spectra directly. 

\begin{table*}
\begin{center}
\begin{small}
\begin{tabular}{l l c c c}
\toprule
{\emph{property}}                  &                & LDA       & QSGW      & experiment\\
\midrule
magnetic moment - Fe$_a$ ($\mu_B$) & $\mu_{a}$    & 3.62      & 4.17      & 4.01 ($R\overline{3}$)~\cite{rodic_true_1999} \\
                                   &                &           &           & 4.11 ($Ia\overline{3}d$)~\cite{rodic_true_1999} \\
magnetic moment - Fe$_d$ ($\mu_B$) & $\mu_{d}$    & 3.48      & 3.93      & 3.95 ($R\overline{3}$)~\cite{rodic_true_1999} \\
                                   &                &           &           & 5.37 ($Ia\overline{3}d$)~\cite{rodic_true_1999} \\
saturation magnetisation (Gauss)   & $4\pi M_s$     &\multicolumn{2}{c}{2459}& 2470~\cite{hansen_saturation_1974}, 3000+~\cite{Gallagher_exceptionally_2016}\\
magnon splitting (meV)             & $\Delta E_m$   & 44        & 23        & 33~\cite{nambu_observation_2020}  \\
spin wave stiffness ($\times10^{-41}$Jm$^{2}$)      & $D$            & 152 & 99 & 83 -- 109~\cite{srivastava_spin_1987}\\
\bottomrule
\end{tabular}
\end{small}
\end{center}
\caption{Summary of low-temperature magnetic properties of YIG.  Calculated magnetic moments are integrated within the respective muffin-tin radii.  We believe that the different moments obtained for the two phases in Ref.~\cite{rodic_true_1999} relate to sensitivity in their fitting procedure and consider their value for Fe$_d$ in the $Ia\overline{3}d$ structure to be unrealistic.\label{tab:comparison}
}
\end{table*}

In studies of YIG it is common to calculate the spectral function $\mathscr{S}^{xy}(\vec{q}, \omega) - \mathscr{S}^{yx}(\vec{q}, \omega)$, which shows the polarisation of the magnons, and we show in Fig.~\ref{fig:magnon_spectrum} for the LDA and QSGW calculations along two lines of symmetry.  Both spectra are similar in overall appearance to each another and to neutron scattering measurements with only some minor differences in the optical modes.  However the energy range the spectrum is spread across is very different in both cases.  

Concerning the QSGW spectrum, the acoustic spectrum (red modes) is very similar to experiments.  The bandwidth of the lowest mode is well reproduced and the energy of the flat bands is also similar to experiments. 
But the optical (blue) modes are much lower in energy, even though they are qualitatively similar to experimental measurements.  
The magnon gap between the parabolic acoustic and optical modes is too small by roughly 80\%.
Overall the QSGW spectrum looks very similar to the previous LDA+U calculations~\cite{xie_first-principles_2017}.  
Here, the purely LDA calculations show optical magnon modes much higher than have been measured and the magnon bandwidth of the lowest mode is too large.  
The magnon gap in the LDA calculation is also larger than the experimental value.

YIG is often treated as a simple ferromagnet for analytical calculations and the spin wave stiffness, $D$, parameterises the dispersion as $\hbar\omega = D q^2$.  This is only valid in the long wavelength (low $q$) limit.  In YIG the dispersion is quadratic in the region $aq <  1$~\cite{harris_spin-wave_1963,cherepanov_saga_1993} (also meaning Bloch's law should be obeyed only to about \SI{40}{K}).  We extract the spin wave stiffness by fitting the quadratic dispersion to the spectrum in this range, finding a value of \SI{99e-41}{Jm^{2}} for the QSGW calculation.  This is very similar to experimental values which vary in the range \SIrange{83e-41}{109e-41}{Jm^{2}} depending on the experimental method used~\cite{srivastava_spin_1987}.  The LDA spin wave stiffness is much higher, \SI{152e-41}{Jm^{2}} as can be seen clearly in the spectrum of Fig.~\ref{fig:magnon_spectrum}.  The QSGW description of the acoustic mode in particular is in very good agreement with neutron scattering studies unlike that of the LDA.

\subsection{Finite temperature magnetisation}

\begin{figure}
\includegraphics[width=0.495\textwidth]{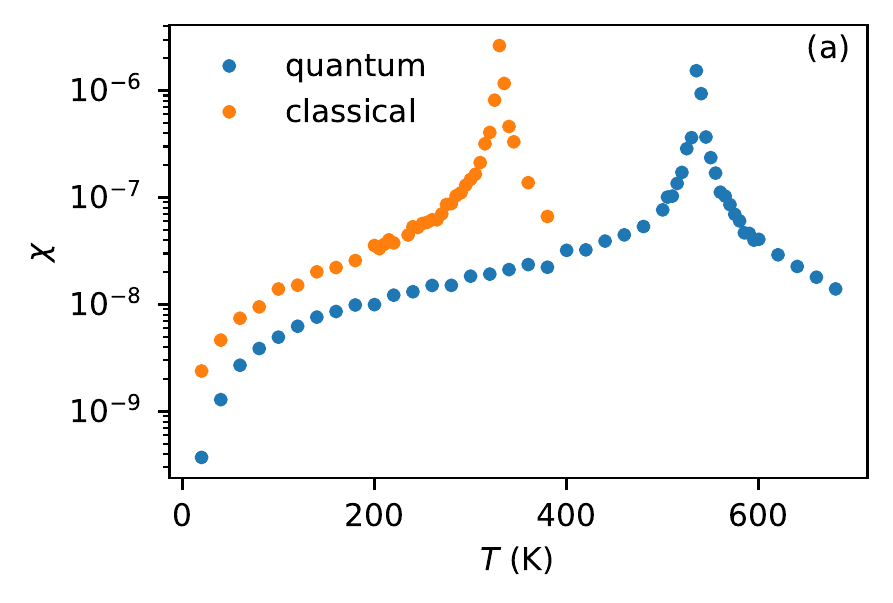}
\includegraphics[width=0.48\textwidth]{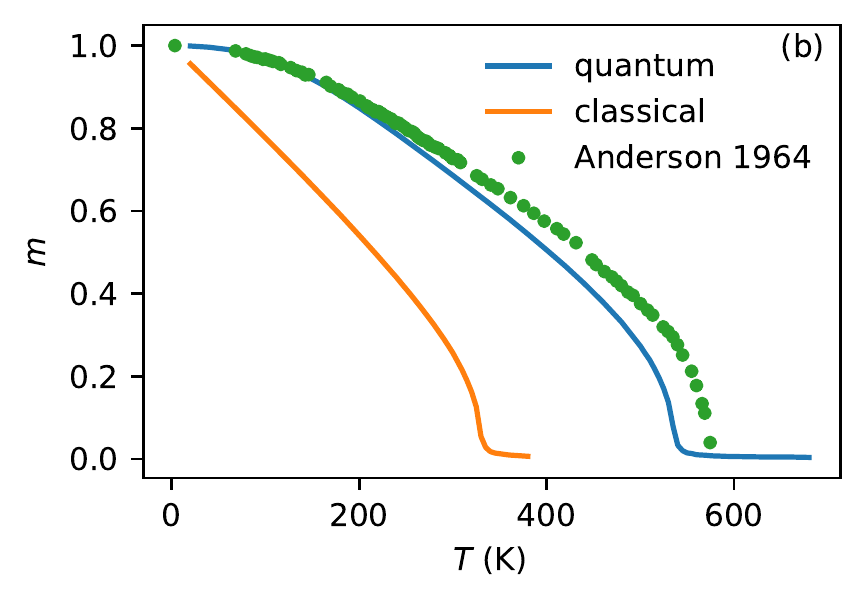}
\caption{Results of the QSGW derived spin model: (a) susceptibility and (b) magnetisation as a function of temperature calculated from spin dynamics with classical (orange) and quantum (blue) thermal statistics. Experimental data in green is from reference~\cite{anderson_molecular_1964}.
\label{fig:magnetisation}}
\end{figure}

Our treatment of finite temperature magnetisation involves extending zero-temperature electronic structure calculations with atomistic spin dynamics (ASD).  Table~\ref{tab:finiteT} lists different approximations to the critical temperature of YIG, obtained using the LDA and QSGW exchange parameters.  Mean field approximation (MFA) and Tyablikov's random phase approximation (RPA)~\cite{Tyablikov1967} assume classical statistics and depend only on $J(q)$.  The atomistic spin dynamics critical temperatures are identified by the divergence of the magnetic susceptibility (Figure~\ref{fig:magnetisation}(a)).  In a quantised spin models and also in our classical spin model with quantum statistics, the Curie temperature depends on both the exchange and the size of the magnetic moments (or the spin $\mathcal{S}$).

\begin{table*}
\begin{center}
\begin{small}
\begin{tabular}{l c c}
\toprule
{\emph{method}}       & \multicolumn{2}{c}{$T_{\rm C}$~(\SI{}{K})}  \\
        & LDA   &       QSGW       \\
\midrule
classical MFA    & 753 & 508 \\
classical RPA   & 465 & 325 \\
classical ASD  & 530  & 320 \\
quantum ASD  & 830 & 535\\
\midrule
experiment~\cite{anderson_molecular_1964} & \multicolumn{2}{c}{559}\\
\bottomrule
\end{tabular}
\end{small}
\end{center}
\caption{Summary of Curie temperature calculations.
\label{tab:finiteT}
}
\end{table*}

The classical mean-field estimates of the Curie temperature are significantly larger than the respective  classical RPA estimates.  The RPA yields an estimate close to the result of classical spin dynamics, which we consider an essentially exact solution of the classical problem, but this result is drastically smaller than the experimental $T_{\rm C}$.  The use of quantum statistics increases the $T_{\rm C}$ estimate considerably, by close to a factor of $2$ (see supplementary information in Ref.~\cite{barker_semiquantum_2019} for $\mathcal{S}$ dependence of $T_{\rm C}$).  The semi-quantum critical temperature estimates are in much better accord with experiment, especially for QSGW.

The form of the QSGW magnetisation curve is shown in Fig.\ref{fig:magnetisation}(b) for both classical and semi-quantum statistics together with experimental data.  The importance of semi-quantum statistics for the low-temperature magnetisation is obvious; unlike the QSGW-classical model, the QSGW-quantum approach provides a very good description of the magnetisation curve, deviating only at high $T$, and underestimating $T_{\rm C}$.  

We attribute the accuracy of the QSGW critical temperature to the excellent description of the acoustic magnon mode by the GSQW exchange constants, this is also reflected in the close agreement of the QSGW spin wave stiffness with experiment.  At higher temperatures the magnetisation deviates from experimental measurements because the QSGW optical magnon modes are lower in energy than measured experimentally; because Planck statistics are heavily weighted to lower frequencies the the differences in optical frequencies don't have too large an effect on the magnetisation. 

The improvement in simulated $m(T)$ using a quantum thermostat is not surprising: the magnon Debye temperature in YIG is close to \SI{1000}{K} (as is true of many magnetic materials), which means that the use of classical (Rayleigh-Jeans) statistics is not valid ($\hbar\omega \ll k_B T$ is not a good approximation), even up to the Curie point.  Our method uses a quantum thermostat in classical spin dynamics and avoids the considerable expense and difficulty of solving the fully quantum problem (e.g. the sign problem in quantum Monte Carlo).  In contrast, calculating the magnetisation with classical statistics fails to reproduce a Bloch's law type curve at low temperature, fails to reproduce the Curie temperature and so is deficient both qualitatively as well as quantitatively.

The magnon spectrum should be the primary quantity to assess the quality of $J_{ij}$ parameterisations because it is from this that the thermodynamics naturally follow through the correct thermal occupation of spin waves, and because it is fundamentally this quantity which the ab initio formulae (those based on perturbation such as these calculations or those using the magnetic force theorem) for magnetic exchange describe.

\section{Conclusions}

We have demonstrated parameter free multi-scale modelling of YIG from first principles.  The dispersion, spin wave stiffness and magnetic moments are well parameterised and the Curie temperature is accurately predicted. This demonstrates simultaneous agreement of low and high temperature properties. Even though the optical magnon splitting is smaller than in experiments, this is for a well understood reason (the systematic overestimate of gaps in QSGW).  We demonstrate the capability of QSGW to provide a completely parameter free, first principles description of the electronic and magnetic properties of complex materials with reliably high accuracy. Our results for the critical temperature highlight the need to use quantum statistics for calculating thermodynamics and for assessing $J$ principally in terms of the spin wave spectrum.  QSGW also allows accurate calculations including $f$-shells~\cite{chantis_quasiparticle_2007}, opening the possibility to calculate other members of the rare-earth garnet family in the future.  Accurate, predictive first principles modelling can play an important role here, for example in parameterising materials such as gadolinium iron garnet which is extremely difficult to measure with neutron scattering (due to Gd having the largest neutron scattering cross-section).

\section*{Acknowledgements}
We thank Mark van Schilfgaarde for interesting and enjoyable discussions while conducting these calculations. 
J.B. acknowledges support from the Royal Society through a University Research Fellowship.
J.J. acknowledges support under the CCP9 project {\emph{Computational Electronic Structure of Condensed Matter}} (part of the Computational Science Centre for Research Communities (CoSeC)). 
D.P. acknowledges PRACE for awarding him access to Marconi100 hosted by CINECA, Italy.
This work was undertaken on the Marconi100 Tier-0 HPC system at CINECA, Italy; ARC4, part of the High Performance Computing facilities at the University of Leeds, UK (\url{https://arc.leeds.ac.uk}) and STFC Scientific Computing Department's SCARF cluster (\url{https://www.scarf.rl.ac.uk}).

\section*{References}

\bibliographystyle{iopart-num-doi}
\bibliography{els-yig}

\end{document}